\documentclass[twocolumn,aps,superscriptaddress,prl]{revtex4}
\usepackage{graphicx}
\usepackage{times}
\usepackage{amsmath}
\usepackage{amssymb}
\usepackage{units}
\usepackage{stmaryrd} 
\DeclareGraphicsExtensions{.pdf,.png,.eps,.jpg}

\begin{document}
\preprint{0}

\title{Ultrafast photo-doping and effective Fermi-Dirac distribution of the Dirac particles in $\mathbf{Bi_{2}Se_{3}}$}

\author{A. Crepaldi}
\affiliation{Institute of Condensed Matter Physics (ICMP), Ecole Polytechnique
F\'ed\'erale de Lausanne (EPFL), CH-1015 Lausanne,
Switzerland}

\author{B. Ressel}
\affiliation{Elettra - Sincrotrone Trieste, Strada Statale 14 km 163.5 Trieste, Italy}

\author{F. Cilento}
\affiliation{Elettra - Sincrotrone Trieste, Strada Statale 14 km 163.5 Trieste, Italy}

\author{M. Zacchigna}
\affiliation{C.N.R. - I.O.M., Strada Statale 14 km 163.5 Trieste, Italy}

\author{C. Grazioli}
\affiliation{Elettra - Sincrotrone Trieste, Strada Statale 14 km 163.5 Trieste, Italy}
\affiliation{now at University of Nova Gorica, Vipavska 11 c, 5270 Ajdovscina, Slovenia}

\author{H. Berger}
\affiliation{Institute of Condensed Matter Physics (ICMP), Ecole Polytechnique
F\'ed\'erale de Lausanne (EPFL), CH-1015 Lausanne,
Switzerland}

\author{Ph. Bugnon}
\affiliation{Institute of Condensed Matter Physics (ICMP), Ecole Polytechnique
F\'ed\'erale de Lausanne (EPFL), CH-1015 Lausanne,
Switzerland}

\author{K. Kern}\affiliation{Institute of Condensed Matter Physics (ICMP), Ecole Polytechnique
F\'ed\'erale de Lausanne (EPFL), CH-1015 Lausanne,
Switzerland}
\affiliation{Max-Plank-Institut f\"ur Festk\"orperforschung,
D-70569, Stuttgart, Germany}

\author{M. Grioni}
\affiliation{Institute of Condensed Matter Physics (ICMP), Ecole Polytechnique
F\'ed\'erale de Lausanne (EPFL), CH-1015 Lausanne,
Switzerland}

\author{F. Parmigiani}
\affiliation{Elettra - Sincrotrone Trieste, Strada Statale 14 km 163.5 Trieste, Italy}
\affiliation{Universit\`a degli Studi di Trieste - Via A. Valerio 2 Trieste, Italy}

\date{\today}

\begin{abstract}
We exploit time- and angle- resolved photoemission spectroscopy to determine the evolution of the out-of-equilibrium electronic structure of the topological insulator $\mathrm{Bi_{2}Se_{3}}$. The response of the Fermi-Dirac distribution to ultrashort IR laser pulses has been studied by modelling the dynamics of the \emph{hot} electrons after optical excitation. We disentangle a large increase of the effective temperature ($T^{*}$) from a shift of the chemical potential ($\mu^{*}$), which is consequence of the ultrafast photodoping of the conduction band. The relaxation dynamics of $T^{*}$ and $\mu^{*}$ are $k$-independent and these two quantities uniquely define the evolution of the excited charge population. We observe that the energy dependence of the non-equilibrium charge population is solely determined by the analytical form of the effective Fermi-Dirac distribution.
\end{abstract}

\maketitle

\begin{figure*}[!t]
  \centering
  \includegraphics[width=0.8\textwidth]{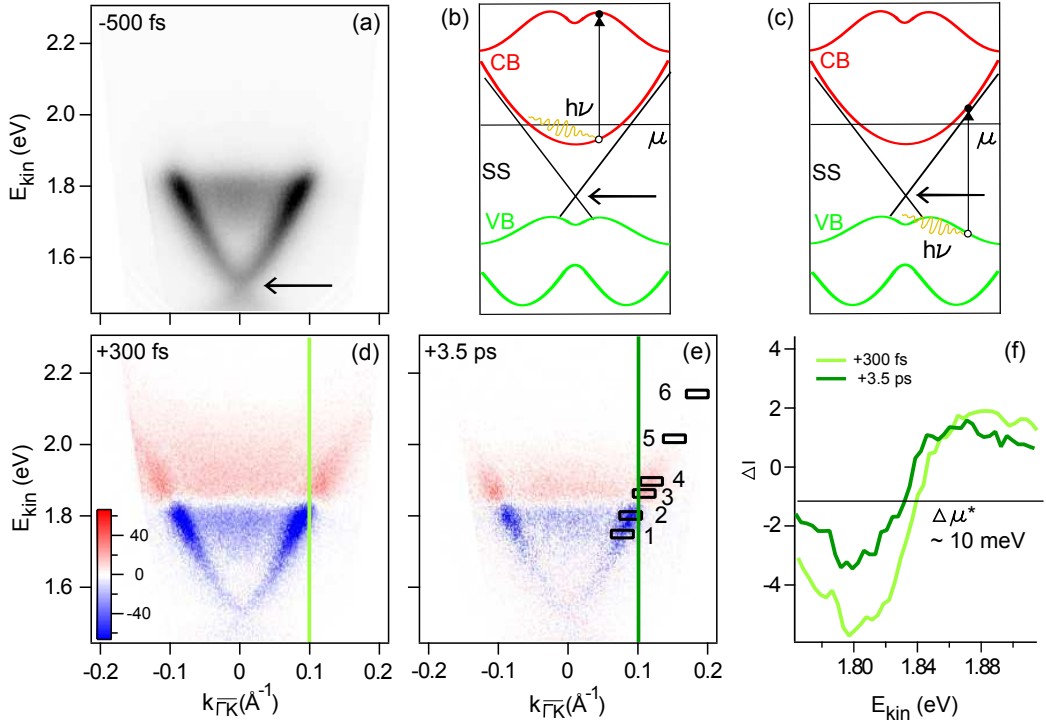}
  \caption[didascalia]{(color online) (a) ARPES band dispersion of $\mathrm{Bi_{2}Se_{3}}$ at negative delay time along the $\overline{\Gamma K}$ high symmetry direction. (b) and (c) schematization of inter-band excitation channels between the conduction-conduction band (CB-CB) (b), and between the valence-conduction bands (VB-CB) (c). (d) and (e) Pump-probe tr-ARPES signal obtained as the difference between the images at positive delays (300 fs and 3.5 ps) and negative delay (-500 fs). Red (blue) represents an increase (decrease) of the spectral weight. (f) Pump-probe difference EDCs at the Fermi wave-vector $k_{F} = 0.1 ~ \mathring{A}^{-1}$, cut from panels (d) and (e). A shift of the chemical potential, $\Delta\mu^{*}$, is experimentally resolved. 
  }
  \label{fig:arpes}
\end{figure*}

The recent discovery of topological insulators (TIs) is renewing the attention on the effects of spin orbit interactions (SOI) in solids, paving the road for the emergence of new quantum states of matter \cite{Hsieh_Nature_2008, Xia_NatPhys_2009, Chen_Science_2009, Chadov_NatMat_2010, Eremeev_NatCom_2012,Bianchi_NatCom_2010, Roushan_Nat_2009, FU_PRL_2007,Zhang_NatPhys_2009}. The SOI acquires particular relevance in the case of systems containing high-\emph{Z} elements, leading to the lifting of the Kramers spin-degeneracy in broken inversion symmetry systems, as described by Rashba \cite{Rashba_1960, Ishi_NatMat_2011, Crepaldi_PRL_2012}, Dresselhaus \cite{Dresselhaus_1955} and Rashba-Bychkov \cite{RB_1984,Lashell_PRL_1996}. Therefore, understanding the consequence of SOI is of primary importance also for future technological applications in spintronics. 

TIs are band insulators (semiconductors) where the conduction and the valence band states have opposite parities and their energy ordering is inverted by the SOI \cite{FU_PRL_2007,Fu_PRB_2007}. The most prominent feature of their electronic structure is the odd number of spin polarized Dirac cones at the surface, connecting the opposite sides of the bulk band gap, resulting in topological protection from backscattering \cite{Roushan_Nat_2009}. Several classes of TIs ($\mathrm{Bi_{x}Sb_{1-x}}$ \cite{Hsieh_Nature_2008}, $\mathrm{Bi_{2}Se_{3}}$ \cite{Xia_NatPhys_2009}, $\mathrm{Bi_{2}Te_{3}}$ \cite{Chen_Science_2009}, $\mathrm{TlBiSe_{2}}$ \cite{Chen_PRL_2010,Kuroda_PRL_2010}, Ternary Heusler compounds \cite{Chadov_NatMat_2010} and Pb based TIs \cite{Eremeev_NatCom_2012,Xu_arxiv_2010}) have been discovered. Among these, $\mathrm{Bi_{2}Se_{3}}$ represents a paradigmatic case, owing to the simplicity of its band structure characterized by a single Dirac cone \cite{Xia_NatPhys_2009, Zhang_NatPhys_2009}. 

In topological insulators, the spin helicity of the metallic surface state offers the unique possibility to support spin-current. Hence, spin-polarized charge distributions can be generated by circularly polarized light \cite{McIver_Nnano_2012, Hsieh_PRL_2011}. The topological protection of the linearly dispersing surface state is expected to strongly affect the scattering mechanisms of the Dirac particle, with respect to normal metallic states. In particular, the different couplings to optical and acoustic phonon modes have been recently studied \cite{Gedik_PRL_2012} and the optical excitation of long-lived electron population in the surface state might play an important role in forthcoming opto-spintronics devices \cite{Gedik_PRL_2012, Sobota_PRL_2012}. 

In this paper we report on the study of the out-of-equilibrium electronic properties of $\mathrm{Bi_{2}Se_{3}}$, investigated by time- and angle- resolved photoemission spectroscopy (tr-ARPES). Although conventional ARPES, with its surface sensitivity, has proven to be effective and rich of information \cite{Hsieh_Nature_2008, Chen_Science_2009, Bianchi_NatCom_2010}, the combined use of ultrashort laser pulses and angle-resolved photoemission experiments offers the unique possibility to explore the dynamical evolution of the charge carriers after an optical excitation, opening the door to the investigation of mechanisms and states hidden to experiments in the frequency domain. After optical excitation, electrons thermalize on a short timescale owing to the fast electron-electron interaction. The resulting \emph{hot} electron gas is described by a Fermi-Dirac distribution with an effective electronic temperature $T^{*}(t)$ and an effective chemical potential $\mu^{*}(t)$ \cite{Allen_PRB_1987,Perfetti_PRL_2007}. From the analysis of the temporal evolution of the Fermi-Dirac function at the Fermi wave-vector $k_{F}$, we succeed in disentangling the large increase of $T^{*}$ from the ultrafast shift in the chemical potential, $\Delta\mu^{*}$.

Our experiment reveals that the dynamics of the non-equilibrium excited electron and hole populations strongly varies with the kinetic energy and the wave-vector $(E,~k)$. We develop a model based on the exponential relaxation of $T^{*}(t)$ and $\mu^{*}(t)$ in the Fermi-Dirac function to quantitatively fit the measured dynamics. We find that the relaxation times of $T^{*}(t)$ and $\mu^{*}(t)$ are $k$-independent. The energy dependence of the charge population is then uniquely determined by the analytical form of the Fermi-Dirac distribution. Our model, owing to the universal properties of the Fermi-Dirac function, is more generally applicable to other tr-ARPES experiments on similar materials.

High quality single crystals of $\mathrm{Bi_{2}Se_{3}}$, in the form of platelets, were grown by the Bridgman technique. The $\mathrm{Bi_{2}Se_{3}}$ samples are n-doped due to atomic vacancies and excess selenium \cite{Xia_NatPhys_2009}. The tr-ARPES experiments were performed at the T-ReX laboratory, Elettra (Trieste), with the use of a commercial Ti:Sa regenerative amplifier (Coherent RegA 9050), producing 800 nm (1.55 eV) laser pulses at 250 kHz repetition rate, with temporal duration of 60 fs. The samples were mounted on a He cryostat held at 100 K, and cleaved \emph{in situ} in ultrahigh vacuum ($2 \cdot 10^{-10}$ mbar) at room temperature. The pump laser light was \emph{p}-polarized, and the absorbed pump fluence was equal to 210 $\mu J/cm^{2}$. The delay between pump and probe was introduced by modifying the optical path of the pump.
Electrons were photoemitted by the \emph{s}-polarized fourth harmonic at 6.2 eV, obtained by harmonic generation in phase-matched BBO crystals. The photo-electrons were analyzed and detected by a SPECS Phoibos 225 hemispherical spectrometer, with energy and angular resolution set respectively equal to 10 meV and $0.3 ^{\circ}$, equivalent to $0.005 ~ \mathring{A}^{-1}$. The overall temporal resolution was set to 300 fs, thus preserving the very high energy resolution required by the present experiment. 

Figure~1~(a) illustrates the band dispersion of $\mathrm{Bi_{2}Se_{3}}$ at negative delay time, \emph{i.e.} before the arrival of the pump pulse (-500 fs), along the $\overline{\Gamma K}$ high symmetry direction. The energy scale reports the measured kinetic energy, since the chemical potential $\mu$ does not represent a good reference for the energy scale in the present experiment, as will be discussed in detail in the following. The linearly-dispersing topological surface state is clearly resolved. A black arrow points towards its apex, \emph{i.e.} the Dirac point. The weaker signal within the Dirac cone at 1.7 - 1.84 eV corresponds to the bottom of the conduction band. These observations are consistent with previous ARPES results from conventional UV sources \cite{Xia_NatPhys_2009, Bianchi_NatCom_2010}. 

The intense pump pulses cause charge excitations between the occupied and unoccupied parts of the conduction band (Fig.~1~(b)), similarly to what observed in metals (CB-CB transitions) \cite{Fann_PRB_1992, Lisowski_AppPhys_2004}. In the present case, the photon energy is larger than the band gap and this enables also inter-band excitations across the gap, from the fully occupied valence band (VB) to the partially unoccupied conduction band (CB) (VB-CB transitions) (Fig.~1~(c)). Figure~1~(d) and (e) show the pump-probe signal obtained as the difference between the ARPES images after and before the optical perturbation. Red (blue) represents an increase (decrease) of the spectral weight. At short delay time (+300 fs) a large density of electrons populates the surface state and the conduction band above the chemical potential \cite{suppl_1}. Such electrons result from intra- and inter- band transitions across the bandgap. However, the temporal resolution in this experiment is not sufficient to capture the formation of the nascent non-equilibrium electron distribution, which is expected to thermalize into a \emph{hot} Fermi-Dirac distribution via electron-electron interaction, with a characteristic time shorter than 40 fs \cite{Park_PRB_2010}. Figure~1~(e) illustrates the reduction in the pump-probe signal at larger delay times (+3.5 ps). The density of excited charge carriers is reduced, as normally observed in metals \cite{Fann_PRB_1992, Lisowski_AppPhys_2004}. 

Despite this similarity, we also observe a deviation from a purely metallic response. Such difference is well captured by a close inspection in the pump-probe difference energy distribution curves (dEDCs) at the Fermi wave-vector $k_{F} = 0.1 ~ \mathring{A}^{-1}$ for various delay times \cite{suppl_1}. Figure~1~(f) shows the two pump-probe dEDCs at +300 fs and +3.5 ps. We focus our attention on the zero-crossing point, which separates the charge-depletion and the charge-accumulation regions. The zero-crossing point clearly shifts more than 10 meV.
We propose that the modification of the zero-crossing energy can be explained by an ultrafast shift of the chemical potential, as recently reported in $\mathrm{Bi_{2}Se_{3}}$ also by Wang et \emph{al.} \cite{Gedik_PRL_2012}. We interpret such an effect as the consequence of the photo-doping resulting from the inter-band excitations across the gap. In the case of a non-degenerate semiconductor (\emph{i.e.} $\mu$ lying in the band gap) the optically excited electrons and holes must be considered as separated systems, each with a distinct thermal distribution, and with a different chemical potential depending on the charge density \cite{Othonos_JApp_1998}. The thermalized electron and hole distributions relax towards equilibrium via radiative processes, Auger recombination or diffusion. A simpler relaxation via diffusion was also proposed for p-doped $\mathrm{Bi_{2}Se_{3}}$ \cite{Sobota_PRL_2012}. The electron-phonon scattering is unsuitable for recombining the excited electrons with holes, because the value of the gap is larger than the highest phonon energy (23 meV) \cite{Richter_1977, Sobota_PRL_2012}. Furthermore, for n-doped $\mathrm{Bi_{2}Se_{3}}$, the band gap is acting as a bottleneck slowing down the relaxation of the photo-excited holes at high binding energies (in VB) and of the low energy holes (in CB). This results in a transient excess of charge carriers in the conduction band, which is at the origin of the proposed ultrafast shift of the chemical potential.

\begin{figure}[!t]
  \centering
  \includegraphics[width=0.475\textwidth]{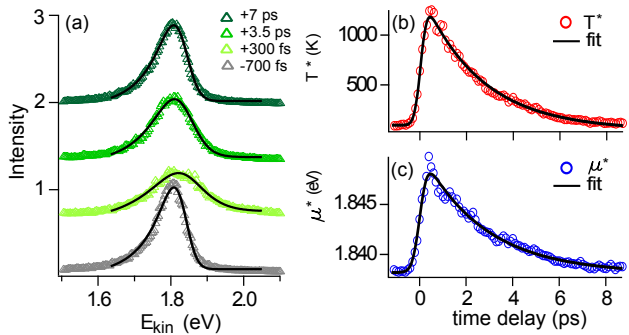}
  \caption[didascalia]{(color on line) (a) EDCs at the Fermi wave-vector $k_{F} = 0.1 ~ \mathring{A}^{-1}$ at four different delay times (-700 fs, +300 fs, +3.5 ps and +7 ps). The fitting curves, in black lines, contain the FD distribution, multiplied with a lorentzian which accounts for the surface state peak. The temporal evolution of the effective temperature $T^{*}$ and of the effective chemical potential $\mu^{*}$, as obtained from the the fit, are shown respectively in panel (b) and (c). A single decay exponential function well fits the temporal dynamics, with characteristic relaxation time equal to $\tau_{T} \sim 2.5 ~ ps$ and $\tau_{\mu} \sim 2.7 ~ ps$, respectively.      
  }
  \label{fig:EDC}
\end{figure}

The emergence of the ultrafast chemical shift is further supported by a quantitative analysis of the temporal evolution of the FD distribution function at $k_{F}$. Figure~2~(a) displays the EDCs at some selected delay times (-700 fs, +300 fs, +3.5 ps and +7 ps). The product of the FD distribution with a lorentzian, describing the surface state peak, is used as model function to fit the spectra \cite{suppl_1}. The values of the effective temperature $T^{*}$ and of the effective chemical potential $\mu^{*}$, as obtained from the \emph{hot} FD, are shown as a function of the delay time in Figure~2~(b) and (c) respectively. Their temporal evolution is fitted with a single decay exponential, with a proper rise-time (modelled by a step-function), convoluted with a gaussian to account for the temporal resolution. The characteristic relaxation times are $\tau_{T} \sim 2.5 ~ ps$ and $\tau_{\mu} \sim 2.7 ~ ps$. The $T^{*}$ relaxation is slightly slower than previously reported for p-doped $\mathrm{Bi_{2}Se_{3}}$ ($1.67 ~ ps$) \cite{Sobota_PRL_2012}. This discrepancy might be ascribed to the different doping, but it could also be related to the larger pump fluence used in the present case. The maximum value of $\Delta\mu^{*}$ is $~ 12~ meV$, in agreement with the pump-probe dEDCs data. 

The creation of a long-lived population in the conduction band has been proposed to be the key-mechanism responsible for the slow relaxation time in the topological insulators $\mathrm{Bi_{2}Se_{3}}$ \cite{Sobota_PRL_2012} and $\mathrm{Bi_{2}Te_{3}}$ \cite{Perfetti_arxiv_2012}. In these models, the conduction band acts as a charge reservoir for the topological surface state at lower binding energies. Such an effect was inferred from the difference in the temporal evolution of the tr-ARPES intensity at different positions in the band structure, $I(E,k,t)$. Also in our investigation, the evolution of the surface state population at different binding energies along its linear dispersion shows a peculiar energy dependent relaxation. However, a detailed analysis of the data suggests a different interpretation.

\begin{figure*}[!t]
  \centering
  \includegraphics[width=0.9\textwidth]{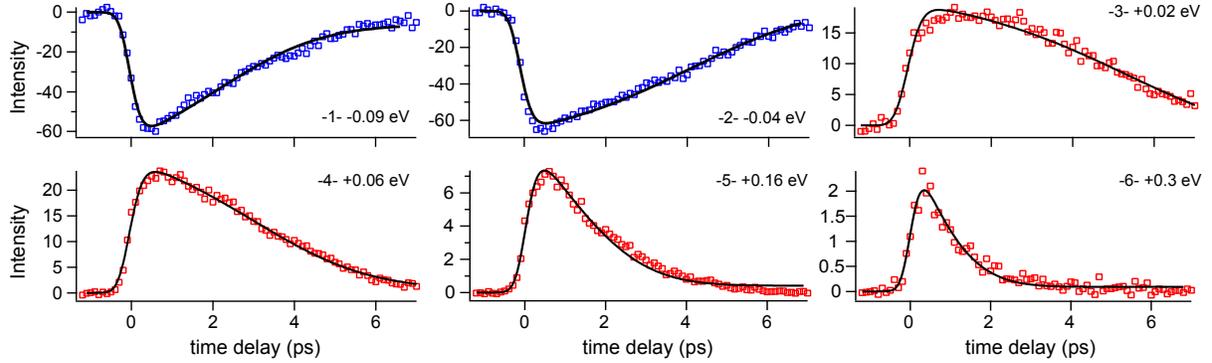}
  \caption[didascalia]{(color online) Ultrafast evolution of the non-equilibrium charge population in the topological surface state at six energies, along its linear dispersion, as indicated in Figure~1~(e).
The results of the fitting routine are displayed as black lines. The model function results from the intensity of the Fermi Dirac distribution at that energy at an effective temperature $T^{*}$ and with an effective chemical potential $\mu^{*}$, which relax exponentially in time as shown respectively in Figure~2~(b) and (c).}
  \label{fig:dynamics}
\end{figure*}

Figure~3 displays the $I(E,k,t)$ signal, for six selected $(E,k)$ positions below and above the chemical potential, as indicated in Figure~1~(e). We focus on the topological surface state and on the evolution of its out-of-equilibrium population. Similar trends are observed in the conduction band, as well. We average the intensity over a narrow energy window (less than 20 meV), and this provides us with a precise information on the \emph{energy dependent} temporal evolution of the excited electrons. This constitutes an important difference to a previous study on $\mathrm{Bi_{2}Te_{3}}$, in which large energy windows were selected, enclosing distinct bands \cite{Perfetti_arxiv_2012}. The dynamics is fast for  $E >> \mu$ (-6-), and \emph{apparently} slows down when approaching $\mu$ (-4- and -3-). The behavior is opposite in sign but symmetric around $\mu$ (-1- and -2-). A similar observation was reported in previous studies of p-doped $\mathrm{Bi_{2}Se_{3}}$, but the dynamics was analyzed only at a qualitative level \cite{Sobota_PRL_2012}. In the following, a simple but effective model is proposed to account for this peculiar energy dependence of $I(E,k,t)$. The spectra at the various kinetic energies are fit with the model function $\zeta(E, t)$, given by the following relation:
\begin{widetext}
\begin{equation}
\zeta(E, t) = \int\limits_{-\infty}^{+\infty}  \Theta(t'-t_{0}) \cdot \frac{C}{1+ exp(E-\mu^{*}(t')/k_{B}T^{*}(t'))}  \cdot G(t-t') dt'  + \phi        .
\end{equation}
\end{widetext}
At time \emph{t} the intensity $\zeta(E, t)$ is the result of the convolution of a gaussian \emph{G}, describing the temporal resolution, with a Fermi-Dirac distribution. The latter is defined by a time-dependent effective temperature $T^{*}(t)$ and an effective chemical potential $\mu^{*}(t)$, which relax in time  as discussed previously. $\Theta$ is a step function which accounts for the rise-time. The intensity $C$ has a weak dependence on the kinetic energy \emph{E}, which we attributed to matrix element effect. $C$ accounts for the variation in sign in the signal between accumulation and depletion, when moving from above to below the chemical potential. Finally, $\phi$ represents a background. In our model the relaxation time is uniquely determined by $\tau_{T}$ and $\tau_{\mu}$, and the apparent energy dependence is conveyed by the Fermi-Dirac distribution. The $\zeta(E, t)$ function is used to fit all the spectra (from -1- to -6-). The results of the fit are reported as black lines in Figure~3, showing a remarkably good agreement with the experimental data.
  
A microscopic description of the temporal evolution of the charge population is complex, since several de-excitation mechanisms must be considered, especially for the recombination of the photo-carriers across the band gap. Previous tr-ARPES study on p-doped $\mathrm{Bi_{2}Se_{3}}$ \cite{Sobota_PRL_2012} and $\mathrm{Bi_{2}Te_{3}}$  \cite{Perfetti_arxiv_2012} analyzed in detail the time evolution of the tr-ARPES intensity $I(E,k,t)$. In these studies a delay between the dynamics of the surface state and the conduction band charge carriers was revealed. This delay was interpreted as the result of a multi-steps relaxation process of the electrons excited in upper branches of the conduction band. Therefore, the conduction band was proposed to act as a charge reservoir for a long-lived electron population in the Dirac cone. The Figure~3 data analysis proves that the long lasting electronic dynamics in the photodoped $\mathrm{Bi_{2}Se_{3}}$ surface state is governed by the evolution of the Fermi-Dirac distribution. Recently, also Wang et \emph{al.} focused their attention on the time evolution of the out-of-equilibrium Fermi-Dirac distribution, in $\mathrm{Bi_{2}Se_{3}}$ \cite{Gedik_PRL_2012}. The authors, by studying as a function of the temperature the response of $\mathrm{Bi_{2}Se_{3}}$ with varied stoichiometries, were able to disentangle the role played by optical and acoustic phonons in the scattering mechanisms between the conduction band and the surface state. From the evolution of $T^{*}$ and $\mu^{*}$ they pointed out that at low temperature the coupling between the bulk and the surface state is suppressed. However, in the present experiment, because of the temperature (100 K), it was not possible to separate the acoustic and optical phonon cooling. Nevertheless, the scope of our work is to interpret the temporal evolution of the tr-ARPES intensity $I(E,k,t)$. This is properly modelled by a time-dependent effective Fermi-Dirac function, uniquely determined by the effective electronic temperature $T^{*}$ and the effective chemical potential $\mu^{*}$. Moreover, our findings indicate that this pump fluence, and the consequent large increase of the effective electronic temperature, might hinder the formation of light-induced spin currents at the Dirac cone. This suggests that the low fluence regime must be investigated in detail in order to reveal some of the non-trivial aspects of the carrier dynamics. 

In summary, we investigated in detail the time-dependent electronic thermal distribution after optical excitation in $\mathrm{Bi_{2}Se_{3}}$. We find that, at all the delay times, the electrons in the surface states are thermalized. Their distribution can be approximated by an effective Fermi-Dirac function, whose temperature $T^{*}$ and chemical potential $\mu^{*}$ relax in time with a decaying exponential behavior. The fast electronic thermalization, below the experimental resolution (300 fs), is ensured by the characteristic time of the electron-electron interaction ($< 40$ fs  \cite{Park_PRB_2010}). The variation in the effective chemical potential is interpreted as the result of the ultrafast photodoping of the conduction band, similarly to what recently reported by Wang et \emph{al.} \cite{Gedik_PRL_2012}. The relaxation time of $T^{*}$ and $\mu^{*}$ are respectively equal to $\tau_{T} \sim 2.5 ~ ps$ and $\tau_{\mu} \sim 2.7 ~ ps$. The former is compatible with a mechanism of energy relaxation to the lattice mediated by the electron-phonon interaction. The latter is related to the relaxation of the excess charge in the conduction band via diffusion. 

We propose a model which relies only on the inherent properties of the Fermi-Dirac distribution function, therefore it provides a viable interpretation for similar energy dependence of the electronic dynamics in a different material \cite{Perfetti_arxiv_2012} or in p-doped $\mathrm{Bi_{2}Se_{3}}$ \cite{Sobota_PRL_2012}. The effect of the topological protection during the relaxation of the Dirac cone charge population may provide the clue for designing new opto-spintronic devices \cite{McIver_Nnano_2012}. Our model paves the way for future investigations between the response of topologically protected and conventional semiconductors.  
    
We thank C. Tournier-Colletta for clarifying discussions. This work was supported in part by the Italian Ministry of University and Research under Grant Nos. FIRB- RBAP045JF2 and FIRB-RBAP06AWK3 and by the European Community\textendash{}Research Infrastructure Action under the FP6 \textquotedblleft{}Structuring the European Research Area\textquotedblright{} Programme through the Integrated Infrastructure Initiative \textquotedblleft{}Integrating Activity on Synchrotron and Free Electron Laser Science\textquotedblright{} Contract No. RII3-CT-2004-506008.


\end{document}